\documentclass[letterpaper, 10 pt, conference]{ieeeconf}  

\IEEEoverridecommandlockouts                              
\usepackage{amsmath}
\usepackage{amssymb}
\usepackage{amsfonts}
\usepackage{mathtools}
\usepackage{algorithm}
\usepackage{algorithmic}
\usepackage{bm}
\usepackage{graphicx}
\graphicspath{{Figures/}}
\usepackage{booktabs}
\usepackage{multirow}
\usepackage{tabularx}

\usepackage{amsthm}
\usepackage{dsfont}
\usepackage{url}

\DeclareMathOperator*{\argmax}{arg\,max}
\newcommand{\R}{\mathbb{R}}
\newcommand{\E}{\mathbb{E}}
\newcommand{\Ind}{\mathds{1}}

\title{\LARGE \bf
Ranking-Augmented On-Policy Optimization with\\
Adaptive Advantage-Normalization for Constrained Control
}

\author{Md Ragib Rownak\textsuperscript{1},
        Sidra Ghayour Bhatti\textsuperscript{1},
        and Qadeer Ahmed\textsuperscript{1}%
\thanks{\textsuperscript{1}Center for Automotive Research,
  The Ohio State University, 930 Kinnear Rd, Columbus, OH 43212, USA.
  E-mail: \texttt{\{rownak.1, bhatti.39, ahmed.358\}@osu.edu}}%
\thanks{Anthropic's Claude~\cite{c_anthropic2025} is used to assist
  with manuscript editing and exposition. All technical content,
  algorithm design, proofs, and experimental results are the sole
  work of the authors.}%
}

\makeatletter
\def\ps@arxivcopyright{%
  \def\@oddhead{}\def\@evenhead{}%
  \def\@oddfoot{\hfil\parbox[t]{\textwidth}{\centering\scriptsize
    \copyright~2026 IEEE. Personal use of this material is permitted.
    Permission from IEEE must be obtained for all other uses, in any
    current or future media, including reprinting/republishing this
    material for advertising or promotional purposes, creating new
    collective works, for resale or redistribution to servers or lists,
    or reuse of any copyrighted component of this work in other works.
    Accepted for presentation at the 2026 IEEE Conference on Decision
    and Control (CDC).}\hfil}%
  \def\@evenfoot{\@oddfoot}}
\makeatother

\begin{document}

\maketitle
\thispagestyle{arxivcopyright}
\pagestyle{empty}

\begin{abstract}

This paper analyzes the boundedness and feasibility properties of
Advantage-Ranked Group Relative Policy Optimization (A-GRPO), a
ranking-augmented, critic-free policy gradient method employing a
Transformer-encoder actor for fixed-horizon
control with terminal constraints. When feasibility is evaluated only
at the final step, the resulting sparse feedback destabilizes
critic-based advantage estimation and weakens standard Lagrangian
approaches. A
trajectory-level ranking mechanism that augments group-relative policy
updates by reweighting advantages according to constraint satisfaction
is formalized, and three results are established: (i)~a scale-adaptive per-timestep normalization bounds
advantage variance at every timestep independently, (ii)~the ranked
advantage strictly separates feasible from violating trajectories
under a verifiable ranking-weight condition, biasing the policy
gradient toward constraint satisfaction, and (iii)~the adaptive dual
variables remain bounded and exhibit a drift-balance property
that acts as a feedback mechanism for feasibility.
These results are validated on a 3,605-step series-hybrid powertrain
energy management task with a terminal state-of-charge constraint,
where A-GRPO achieves $75.4\%$ mean sustained
feasibility with return within $3.7\%$ of the dynamic programming
optimum, outperforming a Proximal Policy Optimization with Lagrangian
penalties (PPO-Lag) baseline ($27.4\%$ sustained), and ablation
experiments confirm that both the ranking and Lagrangian components
are necessary for this performance.

\end{abstract}

\section{INTRODUCTION}

Many control systems operate over a fixed horizon and must satisfy a
constraint evaluated only at the terminal step, such as a
battery state-of-charge (SOC) target at trip
end~\cite{c_egan2023,c_rolando2024}, terminal orbit insertion, or
end-of-run purity requirements. Feasibility is determined by a single
evaluation at step~$T$, making credit assignment across thousands of
preceding steps fundamentally difficult.

Constrained Markov decision process (CMDP) formulations address
feasibility through expected cumulative
cost bounds~\cite{c_altman1999,c_achiam2017}.
Primal-dual methods converge at sublinear rates under
regularity~\cite{c_ding2022,c_paternain2019}, and extensions
including PID-Lagrangian~\cite{c_stooke2020},
barrier~\cite{c_chow2017},
Lyapunov~\cite{c_berkenkamp2017},
reward-constrained~\cite{c_tessler2019}, and
recovery~\cite{c_thananjeyan2021} approaches improve
stability~\cite{c_garcia2015}. However, all
of these methods assume a dense per-step constraint cost.
When the constraint is evaluated only at the terminal step,
the expected-cost formulation collapses to a single scalar
bound on the mean terminal violation.

Proximal Policy Optimization with Lagrangian penalties
(PPO-Lag)~\cite{c_schulman2017,c_ray2019,c_jayant2022} is a common
baseline, but its critic must estimate values over a long
undiscounted horizon ($\gamma = 1$), where bootstrapping errors
compound. Off-policy methods
(SAC~\cite{c_haarnoja2018}, TD3~\cite{c_fujimoto2018}) require
multiple networks; pairing each with a sequence
encoder~\cite{c_agarwal2023,c_yuan2024} is prohibitive for
horizons of thousands of steps. Critic-free methods are therefore
attractive for sequence-model policies. A critic must see the same
observation history as the actor to be consistent with it, so
evaluating a Transformer-encoder policy~\cite{c_vaswani2017} means
training a second network of comparable size at every update.
Group-relative estimation avoids this: the baseline is the
empirical mean return of the rollouts already collected, so the
Transformer actor is the only network trained, and the terminal
penalty enters the advantage exactly through the return-to-go
rather than through a learned value estimate.

Group Relative Policy Optimization (GRPO)~\cite{c_shao2024} replaces
critic-based advantages with group-wise evaluation of rollouts,
avoiding value function instability. Since DeepSeek-R1~\cite{c_guo2025},
variants have addressed acceleration~\cite{c_lin2025},
adaptive objectives~\cite{c_li2025,c_guo2025b}, and continuous
control~\cite{c_khanda2025,c_sane2025}.
However, no existing GRPO variant addresses
terminal constraints, and convergence properties under constrained
settings remain unanalyzed. Learning-based powertrain energy
management~\cite{c_egan2023,c_rolando2024,c_chen2019pt,c_rownak2026} has likewise
not provided formal feasibility guarantees for terminal SOC
constraints.

This paper presents Advantage-Ranked GRPO (A-GRPO), a ranking-augmented,
critic-free policy gradient method for fixed-horizon control with
terminal constraints, and provides a theoretical analysis of its
boundedness and feasibility properties. The contributions are:
\begin{enumerate}
\item A scale-adaptive per-timestep advantage normalization that
  bounds advantage variance at every timestep independently
  (Proposition~1), preventing the gradient collapse that occurs
  under global normalization in long-horizon settings.
\item An advantage-separation property (Theorem~1) showing that the
  A-GRPO ranked advantage systematically favors
  constraint-satisfying trajectories under a verifiable condition
  on the ranking weight.
\item A boundedness and drift-balance result (Theorem~2) for the
  adaptive dual variables ($K_{\mathrm{term}}$ and
  $\lambda_{\mathrm{term}}$), showing that they form a feedback
  mechanism for the chance constraint.
\end{enumerate}

The remainder of this paper is organized as follows.
Section~II reviews relevant background on policy optimization and
constrained MDPs. Section~III formulates the fixed-horizon
constrained control problem. Section~IV presents the A-GRPO method
and its training procedure. Section~V provides the boundedness and
feasibility analysis. Section~VI validates the method on a
terminally constrained powertrain control task, and
Section~VII concludes.

\section{PRELIMINARIES}

\subsection{Finite-Horizon Markov Decision Processes}

A finite-horizon Markov decision process (MDP) is defined by the
tuple $(\mathcal{S}, \mathcal{A}, f, r, T, \gamma)$, where
$\mathcal{S}$ is the state space, $\mathcal{A}$ is the action space,
$f : \mathcal{S} \times \mathcal{A} \to \mathcal{S}$ is the
transition function, $r : \mathcal{S} \times \mathcal{A} \to \R$ is
the reward function, $T$ is the horizon length, and
$\gamma \in (0,1]$ is the discount
factor~\cite{c_sutton2018,c_bertsekas1999}. A stochastic policy
$\pi_\theta(\cdot | o_t)$ parameterized by~$\theta$ maps observations
to action distributions. The undiscounted ($\gamma = 1$) return of a
trajectory $\tau = (o_0, a_0, r_0, \ldots, o_{T-1}, a_{T-1}, r_{T-1})$
is $R(\tau) = \sum_{t=0}^{T-1} r_t$. The objective is to find
$\theta^{\ast} = \argmax_\theta \, \E_{\pi_\theta}[R(\tau)]$.

\subsection{Policy Gradient Methods}

Policy gradient methods optimize $\theta$ by estimating the gradient of
the expected return with respect to the policy
parameters~\cite{c_williams1992,c_sutton2000}. The REINFORCE
estimator takes the form
\begin{equation}\label{eq:reinforce}
  \nabla_\theta J(\theta) = \E_{\pi_\theta}\!\bigg[
    \sum_{t=0}^{T-1} \nabla_\theta \log\pi_\theta(a_t|o_t)\,
    \hat{A}_t\bigg],
\end{equation}
where $\hat{A}_t$ is an advantage estimate that reduces variance
relative to the raw return.
Proximal Policy Optimization (PPO)~\cite{c_schulman2017} stabilizes
updates by maximizing a clipped surrogate objective:
\begin{equation}\label{eq:ppo_prelim}
  L_{\mathrm{clip}}(\theta) = \E\!\Big[\min\!\big(
    \rho_t\,\hat{A}_t,\;
    \mathrm{clip}(\rho_t, 1{-}\epsilon, 1{+}\epsilon)\,\hat{A}_t
  \big)\Big],
\end{equation}
where $\rho_t = \pi_\theta(a_t|o_t)/\pi_{\theta_{\mathrm{old}}}(a_t|o_t)$
is the importance sampling ratio and $\epsilon$ restricts the
step size to prevent destabilizing updates.

\subsection{Group Relative Policy Optimization}

GRPO~\cite{c_shao2024} eliminates the learned value function by
constructing advantages from a group of $B$ parallel rollouts
collected under the current policy. For each rollout~$i$, the
return $R_i = \sum_t r_{t,i}$ is computed, and advantages are
obtained via group-level normalization:
\begin{equation}\label{eq:grpo_adv}
  \hat{A}_i = \frac{R_i - \mu}{\sigma}\,,
\end{equation}
where $\mu$ and $\sigma$ are the mean and standard deviation of
$\{R_i\}_{i=1}^{B}$.
A KL penalty toward a reference policy is typically added to
prevent excessive divergence from the rollout distribution.

\subsection{Constrained MDPs and Lagrangian Relaxation}

A constrained MDP (CMDP)~\cite{c_altman1999} augments the standard
MDP with cost functions $c_k$ and upper bounds $d_k$, yielding the
optimization problem
\begin{equation}\label{eq:cmdp}
  \max_\theta \; \E_{\pi_\theta}[R(\tau)]
  \quad \text{s.t.} \quad
  \E_{\pi_\theta}\!\bigg[\sum_t c_{k,t}\bigg] \leq d_k,
  \;\; \forall\, k.
\end{equation}
The Lagrangian relaxation introduces dual variables
$\lambda_k \geq 0$ and solves the saddle-point problem
$\max_\theta \min_{\lambda \geq 0} \, \E[R(\tau)] -
\sum_k \lambda_k(\E[\sum_t c_{k,t}] - d_k)$.
Primal-dual algorithms alternate between policy updates and
dual variable ascent steps~\cite{c_ding2022,c_paternain2019}.
Under Slater-type regularity, strong duality
holds~\cite{c_paternain2019} and the approach converges at a
sublinear rate.

\section{PROBLEM FORMULATION}

\subsection{Fixed-Horizon MDP with Terminal Constraint}

An episode has a fixed horizon of $T$ steps~\cite{c_sutton2018}. At step
$t \in \{0,\ldots,T{-}1\}$, the agent observes
$o_t = h(x_t, w_t) \in \R^{d_o}$ and
selects $a_t \sim \pi_\theta(\cdot | o_t)$ from a mixed
continuous-discrete space $\mathcal{A}$. The dynamics
$x_{t+1} = f(x_t, a_t, w_t)$ are deterministic with exogenous
input $w_t$. A stage cost $c_t \geq 0$ yields reward $r_t = -c_t$,
and $\gamma = 1$ is used (undiscounted, natural for fixed-horizon problems).

A scalar terminal violation is defined as
\begin{equation}\label{eq:xi_term}
  \xi_{\mathrm{term}} = \max\!\big(|x_T^{\ast} - x_{\mathrm{ref}}| - \varepsilon, \; 0\big),
\end{equation}
where $x_T^{\ast} \in \R$ denotes the scalar constrained component of the
terminal state (e.g., battery state of charge),
$x_{\mathrm{ref}}$ is the target value,
and $\varepsilon > 0$ is a tolerance band. The episode is feasible
if $\xi_{\mathrm{term}} = 0$ and infeasible otherwise. The binary
feasibility indicator is $I_{\mathrm{safe}} = \Ind[\xi_{\mathrm{term}} = 0]$.

The constrained optimization problem is~\cite{c_altman1999,c_bertsekas1999}
\begin{equation}\label{eq:opt_problem}
  \min_\theta \; \E_{\pi_\theta}\!\bigg[\sum_{t=0}^{T-1} c_t\bigg]
  \quad \text{s.t.} \quad
  \Pr_{\pi_\theta}\!\big(\xi_{\mathrm{term}} = 0\big) \geq p_{\mathrm{safe}},
\end{equation}
where $p_{\mathrm{safe}} \in (0,1]$ is a target feasibility rate.

\subsection{Challenges of Terminal Constraints}

Problem~\eqref{eq:opt_problem} differs from standard
CMDPs~\cite{c_achiam2017} in two respects.
First, sparsity: the constraint is evaluated only at step~$T$,
so Lagrangian approaches must propagate the shaped reward
$\tilde{r}_{T-1} = r_{T-1} - \lambda_{\mathrm{term}}\,\xi_{\mathrm{term}} - \psi_{\mathrm{term}}\,\Ind[\xi_{\mathrm{term}} {>} 0]$ backward
through the full horizon, which is difficult for critic-based methods
when $T$ is large.
Second, batch heterogeneity: a global multiplier~$\lambda_{\mathrm{term}}$
acts on the mean violation
$\bar{\xi} = (1/B)\sum_i \xi_{\mathrm{term},i}$, providing no
information about which rollouts violated and by how much.
These observations motivate a trajectory-level ranking mechanism
that provides a within-batch relative signal complementing the
Lagrangian penalty.

\section{RANKING-AUGMENTED POLICY GRADIENT}

\subsection{Lagrangian Reward Shaping}

The reward stream is augmented using adaptive nonnegative multipliers
$\lambda_{\mathrm{tail}}$ and $\lambda_{\mathrm{term}}$:
\begin{align}
  \tilde{r}_t &= r_t - \lambda_{\mathrm{tail}}\,\xi_{\mathrm{tail},t}
               - \psi_{\mathrm{tail}}\,\Ind[\xi_{\mathrm{tail},t} {>} 0],
               \nonumber\\
              &\qquad t < T{-}1, \label{eq:r_shaped}\\
  \tilde{r}_{T\!-\!1} &= r_{T\!-\!1}
               - \lambda_{\mathrm{term}}\,\xi_{\mathrm{term}}
               - \psi_{\mathrm{term}}\,\Ind[\xi_{\mathrm{term}} {>} 0].
               \label{eq:r_shaped_term}
\end{align}
where
$\xi_{\mathrm{tail},t} = \max(|x_t^{\ast} - x_{\mathrm{ref}}| - \varepsilon,\, 0)$
for steps in the tail region (final portion of the horizon) and
zero otherwise,
and $\psi_{\mathrm{tail}},\psi_{\mathrm{term}} \geq 0$ are fixed
penalties improving feasible/infeasible separability.
The undiscounted return-to-go is
$G_{t,i} = \sum_{k=t}^{T-1} \tilde{r}_{k,i}$.

\subsection{Group-Relative Advantage with Per-Timestep Normalization}

Following GRPO~\cite{c_shao2024}, advantages are constructed from
empirical returns without a critic. Given $B$ parallel rollouts,
returns are normalized at each timestep independently:
\begin{equation}\label{eq:adv_norm}
  \hat{A}_{t,i} = \frac{G_{t,i} - \mu_t}{\max(\sigma_t,\; \varphi_t)}\,,
\end{equation}
where
\begin{equation}\label{eq:mu_sigma}
  \mu_t = \frac{1}{B}\sum_{i=1}^{B} G_{t,i}\,, \quad
  \sigma_t = \sqrt{\frac{1}{B}\sum_{i=1}^{B}(G_{t,i} - \mu_t)^2}\,,
\end{equation}
and $\varphi_t$ is a scale-adaptive floor defined as
\begin{equation}\label{eq:eps_floor}
  \varphi_t = c_\varphi \cdot \frac{1}{B}\sum_{i=1}^{B}|G_{t,i}| + \nu\,,
\end{equation}
with $c_\varphi > 0$, $\nu > 0$. The floor prevents advantage explosion
when $\sigma_t$ collapses at convergence, while adapting to the
return magnitude. Properties are analyzed in Section~V.

\subsection{Terminal Feasibility Ranking}

For trajectory-level differentiation beyond the scalar Lagrangian,
the advantage is augmented with two ranking terms.
The hinge signal is
\begin{equation}\label{eq:hinge}
  \bar{\xi}_i = \mathrm{clip}\!\left(\frac{\xi_{\mathrm{term},i}}{\varepsilon},\; -\bar{c},\; \bar{c}\right),
\end{equation}
where $\varepsilon$ is the tolerance band from~\eqref{eq:xi_term} and
$\bar{c} > 0$ is a clipping bound.
The centering signal is
\begin{equation}\label{eq:center}
  \bar{e}_i = \mathrm{clip}\!\left(
    \frac{|x_{T,i}^{\ast} - x_{\mathrm{ref}}|}{\varepsilon},\; 0,\; 1
  \right).
\end{equation}
The ranked advantage is then
\begin{equation}\label{eq:ranked_adv}
  A_{t,i} = \hat{A}_{t,i} - K_{\mathrm{term}}\,\bar{\xi}_i
           - K_{\mathrm{center}}\,\bar{e}_i
           + s\,I_{\mathrm{safe},i}\,,
\end{equation}
where $K_{\mathrm{term}} \geq 0$ controls feasibility enforcement,
$K_{\mathrm{center}} \geq 0$ provides within-band precision, and
$s \geq 0$ is a fixed feasibility shift that rewards
constraint-satisfying rollouts with a constant advantage bonus
($I_{\mathrm{safe},i}$ is the binary feasibility indicator
from~\eqref{eq:xi_term}).
All three ranking terms apply uniformly across timesteps~$t$ for
rollout~$i$, since the terminal constraint depends on cumulative
actions.

\subsection{Adaptive Violation Pressure}

The ranking weight $K_{\mathrm{term}}$ is adapted using the batch
feasibility rate $S_n = (100/B)\sum_i I_{\mathrm{safe},i}$
at update~$n$:
\begin{equation}\label{eq:kterm_update}
  K^{(n+1)} \!=\! \begin{cases}
    \min\!\big(K_{\max},\, K^{(n)} \!\cdot\! \kappa_{\uparrow}\big),
      & \!\text{if } S_n < \tau,\\[2pt]
    \max\!\big(K_{\min},\, K^{(n)} \!\cdot\! \kappa_{\downarrow}\big),
      & \!\text{if } S_n \geq \tau,
  \end{cases}
\end{equation}
where $K^{(n)} \equiv K_{\mathrm{term}}^{(n)}$,
$\tau \equiv \tau_{\mathrm{safe}}$,
$\kappa_{\uparrow} > 1$ and $\kappa_{\downarrow} < 1$ are
multiplicative factors, and $0 \leq K_{\min} \leq K_{\max}$ bound the range.
This implements a feedback mechanism for
enforcing~\eqref{eq:opt_problem}.

\subsection{Lagrangian Multiplier Updates}

After each policy update, the multipliers are adjusted using
batch-averaged violations:
\begin{equation}\label{eq:lambda_update}
  \lambda^{(n+1)} \!=\! \begin{cases}
    \lambda^{(n)} + \alpha_\lambda \, \bar{\xi}^{(n)},
      & \!\text{if } \bar{\xi}^{(n)} > 0,\\[2pt]
    (1 {-} \delta_\lambda)\,\lambda^{(n)},
      & \!\text{if } \bar{\xi}^{(n)} = 0,
  \end{cases}
\end{equation}
with $\lambda \equiv \lambda_{\mathrm{term}}$,
$\bar{\xi}^{(n)} = (1/B)\sum_i \xi_{\mathrm{term},i}^{(n)}$,
step size $\alpha_\lambda > 0$, decay $\delta_\lambda \in (0,1)$,
and projection to $[0, \lambda_{\max}]$.
The $\lambda_{\mathrm{tail}}$ update follows the same structure.

\subsection{Policy Update}

The policy is updated using the clipped surrogate
objective~\cite{c_schulman2017} with the ranked
advantages~\eqref{eq:ranked_adv}:
\begin{multline}\label{eq:ppo_clip}
  L_{\mathrm{clip}}(\theta) = \frac{1}{BT}\sum_{t,i}
    \min\!\big(\rho_{t,i}(\theta)\,A_{t,i},\;\\
    \mathrm{clip}(\rho_{t,i}(\theta), 1{-}\epsilon, 1{+}\epsilon)\,A_{t,i}\big),
\end{multline}
where $\rho_{t,i}(\theta) = \pi_\theta(a_{t,i}|o_{t,i}) / \pi_{\theta_{\mathrm{old}}}(a_{t,i}|o_{t,i})$
is the importance ratio and $\epsilon > 0$ is the clip parameter.
The full loss includes a KL penalty toward a reference
policy~$\pi_{\theta_{\mathrm{ref}}}$ and entropy regularization:
\begin{equation}\label{eq:full_loss}
  J(\theta) = -L_{\mathrm{clip}}(\theta) + \beta_{\mathrm{KL}}\,
  \E[\widehat{\mathrm{KL}}] - \eta_n\,H_{\mathrm{cont}}
  - \eta_\delta\,H_{\mathrm{disc}}\,,
\end{equation}
where $\widehat{\mathrm{KL}}$ is a clipped, nonnegative per-sample
KL estimator:
\begin{align}
  \widehat{\mathrm{KL}}_{t,i} &= \big(e^{\Delta_{t,i}} - 1\big) - \Delta_{t,i}\,, \label{eq:kl_est}\\
  \Delta_{t,i} &= \mathrm{clip}\!\big(\!\log\pi_\theta(a_{t,i}|o_{t,i}) {-} \log\pi_{\theta_{\mathrm{ref}}}(a_{t,i}|o_{t,i}),\; {-}20, 20\big), \nonumber
\end{align}
$H_{\mathrm{cont}}$ and $H_{\mathrm{disc}}$ are the differential
entropies of the continuous and discrete action heads,
$\beta_{\mathrm{KL}} > 0$ is the KL penalty weight,
$\eta_\delta > 0$ is a fixed discrete-entropy coefficient, and
$\eta_n$ is a scheduled continuous-entropy coefficient that
transitions from exploration encouragement to a compression penalty:
\begin{equation}\label{eq:eta_schedule}
  \eta_n = \begin{cases}
    \;\;\,\eta, & n < n_p \quad (\text{exploration}),\\[2pt]
    -\eta_p, & n \geq n_p \quad (\text{precision}),
  \end{cases}
\end{equation}
where $\eta > 0$ and $\eta_p > 0$ are the exploration and precision
coefficients and
$n_p = \lfloor 0.8\,N \rfloor$ is the precision-phase onset.
If the minibatch mean KL exceeds a threshold $\mathrm{KL}_{\mathrm{skip}}$,
that minibatch is discarded entirely (hard trust-region safeguard).

\subsection{Training Procedure}

The complete training loop is summarized in Algorithm~\ref{alg:agrpo}.
Each iteration collects parallel rollouts, constructs ranked
advantages through the three layers formalized above, and updates
both the policy and the dual variables.
The three layers~\eqref{eq:r_shaped}--\eqref{eq:ranked_adv}, together
with the adaptive updates~\eqref{eq:kterm_update}--\eqref{eq:lambda_update},
carry the terminal constraint into the policy gradient, and Section~V
analyzes their normalization, separation, and boundedness properties.
The clipped surrogate, the KL penalty with its minibatch safeguard,
the entropy schedule, and the reference-policy refresh are standard
on-policy components retained from PPO and GRPO practice; they
regulate the size and stability of each policy update. Sensitivity to
the principal design factors of both groups is reported in
Section~VI-C.

\begin{algorithm}[h]
\caption{A-GRPO: Ranking-Augmented Policy Gradient}
\label{alg:agrpo}
\begin{algorithmic}[1]
\REQUIRE Horizon $T$, rollouts $B$, clip $\epsilon$, KL weight $\beta_{\mathrm{KL}}$
\STATE Initialize $\pi_\theta$, $\pi_{\theta_{\mathrm{ref}}} \!\leftarrow\! \pi_\theta$, $\lambda_{\mathrm{tail}}$, $\lambda_{\mathrm{term}}$, $K_{\mathrm{term}}$
\FOR{update $n = 1$ to $N$}
  \STATE Collect $B$ rollouts $\{\tau_i\}_{i=1}^{B}$ of length $T$ under $\pi_\theta$
  \STATE Shape rewards $\tilde{r}_{t,i}$ via~\eqref{eq:r_shaped}--\eqref{eq:r_shaped_term} \hfill \textit{[Lagrangian layer]}
  \STATE Compute returns-to-go $G_{t,i} = \sum_{k=t}^{T-1} \tilde{r}_{k,i}$
  \STATE Normalize $\hat{A}_{t,i}$ per-timestep via~\eqref{eq:adv_norm}--\eqref{eq:eps_floor} \hfill \textit{[Prop.~1]}
  \STATE Rank $A_{t,i}$ via~\eqref{eq:ranked_adv} using~\eqref{eq:hinge}--\eqref{eq:center} \hfill \textit{[Thm.~1]}
  \STATE Update $\theta$ by minimizing~\eqref{eq:full_loss} over minibatches; skip if $\widehat{\mathrm{KL}} > \mathrm{KL}_{\mathrm{skip}}$
  \STATE Adapt $K_{\mathrm{term}}$ via~\eqref{eq:kterm_update}; adapt $\lambda_{\mathrm{term}},\lambda_{\mathrm{tail}}$ via~\eqref{eq:lambda_update} \hfill \textit{[Thm.~2]}
  \STATE Update $\pi_{\theta_{\mathrm{ref}}} \!\leftarrow\! \pi_\theta$ if mean shaped return improves
\ENDFOR
\end{algorithmic}
\end{algorithm}

\section{BOUNDEDNESS AND FEASIBILITY ANALYSIS}

\subsection{Advantage Stability under Per-Timestep Normalization}

Under global normalization, a single mean~$\mu$ and standard
deviation~$\sigma$ are computed over all $(t,i)$ pairs, yielding
$\hat{A}_{t,i} = (G_{t,i} - \mu)/\sigma$.
The per-timestep variance of these advantages is
$\mathrm{Var}_t(\hat{A})
= (1/B)\sum_i (\hat{A}_{t,i} - \bar{\hat{A}}_t)^2
= \sigma_t^2/\sigma^2$,
where the global shift~$\mu$ cancels because variance is
translation-invariant
($\bar{\hat{A}}_t = (\mu_t {-} \mu)/\sigma$ drops out).
Since the return-to-go $G_{t,i} = \sum_{k=t}^{T-1}\tilde{r}_{k,i}$
sums $T {-} t$ reward terms,
$\mathrm{Var}(G_{t,i})$ decreases with~$t$
whenever per-step reward covariances are non-negative
(the typical case when rollouts share a common exogenous input):
early returns-to-go aggregate variance from all future rewards,
while $G_{T-1,i}$ contains only the final reward.
Thus $\sigma_0 > \sigma_{T-1}$ and
$\mathrm{Var}_{T-1}(\hat{A}) < 1$, suppressing the gradient signal
at the terminal step where constraint satisfaction is most needed.
Per-timestep normalization~\eqref{eq:adv_norm} resolves this by
normalizing each timestep independently:

\textbf{Proposition~1.}
Let $\{G_{t,i}\}_{i=1}^{B}$ be returns-to-go at timestep~$t$ across
$B$ rollouts, and let $\hat{A}_{t,i}$ be defined
by~\eqref{eq:adv_norm}--\eqref{eq:eps_floor} with constants
$c_\varphi > 0$, $\nu > 0$. Then:

(a) In the normal regime ($\sigma_t > \varphi_t$):
the empirical variance satisfies $\mathrm{Var}(\hat{A}_t) = 1$
across the batch.

(b) In the collapsed regime ($\sigma_t \leq \varphi_t$):
\begin{equation}\label{eq:adv_bound}
  |\hat{A}_{t,i}| = \frac{|G_{t,i} - \mu_t|}{\varphi_t}
  = \frac{|G_{t,i} - \mu_t|}{c_\varphi\,\bar{G}_t + \nu}\,,
\end{equation}
where $\bar{G}_t = (1/B)\sum_i |G_{t,i}|$.

\textbf{Proof.}
\emph{Part~(a).}
When $\sigma_t > \varphi_t$, the denominator
in~\eqref{eq:adv_norm} equals~$\sigma_t$, so
$\hat{A}_{t,i} = (G_{t,i} - \mu_t)/\sigma_t$.
Since $\mu_t$ and $\sigma_t$ are the empirical mean and standard
deviation of $\{G_{t,i}\}_{i=1}^B$, the normalized values satisfy
$\sum_i \hat{A}_{t,i} = 0$ and
$(1/B)\sum_i \hat{A}_{t,i}^2 = 1$ by construction.
Because the empirical mean is zero, the second moment equals the
variance: $\mathrm{Var}(\hat{A}_t) = (1/B)\sum_i \hat{A}_{t,i}^2 = 1$.

\emph{Part~(b).}
When $\sigma_t \leq \varphi_t$, the denominator
in~\eqref{eq:adv_norm} equals~$\varphi_t$ instead
of~$\sigma_t$. Each advantage is therefore bounded as
$|\hat{A}_{t,i}| = |G_{t,i} - \mu_t|/\varphi_t$.
Substituting the floor definition~\eqref{eq:eps_floor},
$\varphi_t = c_\varphi\,\bar{G}_t + \nu \geq \nu > 0$,
so $|\hat{A}_{t,i}| = |G_{t,i} - \mu_t|/(c_\varphi\,\bar{G}_t + \nu)$.
The bound is finite for all~$t$, and as the policy converges
($\sigma_t \to 0$), the floor scales with the mean return
magnitude~$\bar{G}_t$, preventing the advantage from growing
unboundedly even as returns collapse to a consensus.
\hfill$\square$

\subsection{Ranked-Advantage Separation}

It is now shown that the ranked advantage~\eqref{eq:ranked_adv} biases
the policy gradient toward feasible trajectories.

\textbf{Theorem~1} (Advantage Separation).
Consider a batch of $B$ rollouts at policy update~$n$. Define the
feasible and violating subsets
$\mathcal{F} = \{i : \xi_{\mathrm{term},i} = 0\}$ and
$\mathcal{V} = \{i : \xi_{\mathrm{term},i} > 0\}$, with
$|\mathcal{F}| \geq 1$ and $|\mathcal{V}| \geq 1$.
Since $\xi_{\mathrm{term},i} > 0$ for every $i \in \mathcal{V}$,
the hinge~\eqref{eq:hinge} gives $\bar{\xi}_i > 0$, so the
denominator below is strictly positive.
If the ranking weight satisfies
\begin{equation}\label{eq:kterm_condition}
  K_{\mathrm{term}} >
  \frac{2\,\max_{t,i}|\hat{A}_{t,i}|}
       {\min_{j \in \mathcal{V}} \bar{\xi}_j}\,,
\end{equation}
then for every timestep $t$ (the ranking terms
in~\eqref{eq:ranked_adv} are independent of~$t$):

(i) The mean ranked advantage of violating rollouts is strictly
below that of feasible rollouts:
\begin{equation}\label{eq:adv_separation}
  \frac{1}{|\mathcal{V}|}\!\sum_{i \in \mathcal{V}}\! A_{t,i}
  < \frac{1}{|\mathcal{F}|}\!\sum_{j \in \mathcal{F}}\! A_{t,j}\,.
\end{equation}

(ii) In the clipped surrogate~\eqref{eq:ppo_clip}, feasible
rollouts receive systematically larger advantage coefficients
than violating rollouts in the policy gradient sum.

\textbf{Proof.}
\emph{Part~(i): Advantage separation.}

\emph{Step~1: Feasible rollouts} ($j \in \mathcal{F}$).
Since $\xi_{\mathrm{term},j} = 0$, the
hinge~\eqref{eq:hinge} gives $\bar{\xi}_j = 0$ and
$I_{\mathrm{safe},j} = 1$. The centering
signal~\eqref{eq:center} satisfies $\bar{e}_j \in [0,1]$
(the terminal state may lie anywhere within the feasible band).
Substituting into~\eqref{eq:ranked_adv}:
\begin{equation}\label{eq:A_feas}
A_{t,j} = \hat{A}_{t,j} - K_{\mathrm{center}}\bar{e}_j + s\,.
\end{equation}

\emph{Step~2: Violating rollouts} ($i \in \mathcal{V}$).
Since $\xi_{\mathrm{term},i} > 0$,
equation~\eqref{eq:xi_term} implies
$|x_{T,i}^{\ast} - x_{\mathrm{ref}}| > \varepsilon$. The
centering signal~\eqref{eq:center} has argument
$|x_{T,i}^{\ast} - x_{\mathrm{ref}}|/\varepsilon > 1$, so
the clip forces $\bar{e}_i = 1$. Also $I_{\mathrm{safe},i} = 0$
and $\bar{\xi}_i > 0$. Substituting:
\begin{equation}\label{eq:A_viol}
A_{t,i} = \hat{A}_{t,i} - K_{\mathrm{term}}\bar{\xi}_i
- K_{\mathrm{center}}\,.
\end{equation}

\emph{Step~3: Pairwise comparison.}
For any pair $(j,i)$ with $j \in \mathcal{F}$,
$i \in \mathcal{V}$, subtracting~\eqref{eq:A_viol}
from~\eqref{eq:A_feas}:
\begin{align}
  A_{t,j} - A_{t,i} &= \underbrace{(\hat{A}_{t,j} - \hat{A}_{t,i})}_{\text{normalized return diff.}}
    + \underbrace{K_{\mathrm{term}}\bar{\xi}_i}_{\text{hinge penalty}} \nonumber\\
    &\quad + \underbrace{K_{\mathrm{center}}(1 - \bar{e}_j)}_{\geq\, 0}
    + \underbrace{s\vphantom{(1)}}_{\geq\, 0}\,.
    \label{eq:pair_bound}
\end{align}
Both $K_{\mathrm{center}}(1 - \bar{e}_j) \geq 0$
(since $\bar{e}_j \leq 1$) and $s \geq 0$
favor the feasible rollout. Dropping these non-negative terms
yields
\begin{equation}\label{eq:pair_lower}
  A_{t,j} - A_{t,i} \geq
    (\hat{A}_{t,j} - \hat{A}_{t,i}) + K_{\mathrm{term}}\bar{\xi}_i\,.
\end{equation}
The worst case for the return difference term is
when it is maximally negative. By the triangle inequality,
\begin{equation}\label{eq:triangle}
|\hat{A}_{t,j} - \hat{A}_{t,i}|
\leq |\hat{A}_{t,j}| + |\hat{A}_{t,i}|
\leq 2\max_{\ell,k}|\hat{A}_{\ell,k}|\,,
\end{equation}
so $\hat{A}_{t,j} - \hat{A}_{t,i} \geq
-2\max_{\ell,k}|\hat{A}_{\ell,k}|$.
Substituting into~\eqref{eq:pair_lower} and using
$\bar{\xi}_i \geq \min_{m \in \mathcal{V}} \bar{\xi}_m$:
\[
A_{t,j} - A_{t,i} \geq
K_{\mathrm{term}} \min_{m \in \mathcal{V}} \bar{\xi}_m
- 2\max_{\ell,k}|\hat{A}_{\ell,k}|\,.
\]
Under condition~\eqref{eq:kterm_condition}, the right-hand side
is strictly positive. Since this holds for every
$(j,i) \in \mathcal{F} \times \mathcal{V}$ at every timestep~$t$,
averaging over all such pairs gives
\[
\frac{1}{|\mathcal{F}|\,|\mathcal{V}|}
\sum_{j \in \mathcal{F}}\sum_{i \in \mathcal{V}}
(A_{t,j} - A_{t,i}) > 0\,,
\]
which rearranges to~\eqref{eq:adv_separation}.

\emph{Part~(ii): Gradient direction (approximate).}
Near the current policy, where $\rho_{t,i}(\theta) \approx 1$
and the clip in~\eqref{eq:ppo_clip} is inactive, the surrogate
gradient is approximately~\cite{c_sutton2000,c_williams1992}:
\begin{multline}\label{eq:grad_decomp}
  \nabla_\theta L_{\mathrm{clip}} \approx
  \frac{1}{BT}\sum_{t,i} A_{t,i}\,
  \nabla_\theta \log\pi_\theta(a_{t,i}|o_{t,i}).
\end{multline}
Decomposing into feasible and violating contributions:
$\nabla_\theta L_{\mathrm{clip}} \approx
(1/BT)\big[\sum_{t,j \in \mathcal{F}} A_{t,j}\,g_{t,j}
+ \sum_{t,i \in \mathcal{V}} A_{t,i}\,g_{t,i}\big]$,
where $g_{t,i} = \nabla_\theta \log\pi_\theta(a_{t,i}|o_{t,i})$.
By part~(i), the $\mathcal{F}$-terms carry strictly higher
advantage coefficients than the $\mathcal{V}$-terms at each
timestep. The net directional effect on $\pi_\theta$ also
depends on the alignment of the individual score vectors
$g_{t,i}$ in parameter space; part~(i) guarantees that
feasible terms receive systematically larger coefficients,
but the resulting parameter update direction is
alignment-dependent.
\hfill$\square$

\textbf{Remark~1.}
Condition~\eqref{eq:kterm_condition} is a sufficient condition that
may not hold at every update, particularly in early training when
$K_{\mathrm{term}}$ is small and $|\hat{A}_{t,i}|$ can be large,
or when near-boundary violations make
$\min_{j \in \mathcal{V}} \bar{\xi}_j$ small.
However, the adaptive schedule~\eqref{eq:kterm_update} increases
$K_{\mathrm{term}}$ whenever $S_n < \tau_{\mathrm{safe}}$,
while policy convergence reduces $|\hat{A}_{t,i}|$ over time, so the
condition is progressively easier to satisfy.
For rollouts with large violations ($\bar{\xi}_j$ bounded away
from zero), the condition is met once $K_{\mathrm{term}}$
is sufficiently large.

\textbf{Remark~2.}
When $\mathcal{F} = \emptyset$, Theorem~1 does not apply directly,
since no feasible rollouts are available for comparison.
Nevertheless, the ranking~\eqref{eq:ranked_adv} still provides
useful differentiation: rollouts with smaller violations receive
higher ranked advantages than those with larger violations,
assigning higher surrogate weights to lower-violation rollouts
and providing a relative signal even in the fully infeasible case.
This stands in contrast to the Lagrangian penalty, which applies
a uniform batch-level signal via the mean violation~$\bar{\xi}$
and cannot distinguish among rollouts of differing violation
magnitude.

\subsection{Boundedness and Drift Balance of Adaptive Multipliers}

The multiplier updates can be analyzed individually using tools from
stochastic approximation~\cite{c_borkar2008} and subgradient
methods for constrained optimization~\cite{c_nedic2009}.
Although $K_{\mathrm{term}}$ and $\lambda_{\mathrm{term}}$ are
coupled through the policy, the following properties hold for
each update rule independently.

\textbf{Theorem~2} (Dual Variable Boundedness and Balance).
Consider the update rules~\eqref{eq:kterm_update}
and~\eqref{eq:lambda_update} with $\kappa_\uparrow > 1$,
$\kappa_\downarrow < 1$, $\alpha_\lambda > 0$, $\delta_\lambda \in (0,1)$,
and bounded ranges $K_{\mathrm{term}} \in [K_{\min}, K_{\max}]$,
$\lambda_{\mathrm{term}} \in [0, \lambda_{\max}]$. Then:

(i) Boundedness:
$K_{\mathrm{term}}^{(n)} \in [K_{\min}, K_{\max}]$ and
$\lambda_{\mathrm{term}}^{(n)} \in [0, \lambda_{\max}]$ for all~$n$,
by construction of the clipping operations.

(ii) $\lambda$-relaxation (conditional):
If the batch mean violation satisfies
$\bar{\xi}_{\mathrm{term}}^{(n)} = 0$ for all
$n \geq n_0$ (i.e., all $B$ rollouts are simultaneously feasible
for every update after~$n_0$, a strong condition that is
rarely satisfied in practice; see Section~VI-B), then
$\lambda_{\mathrm{term}}^{(n)} \leq
(1-\delta_\lambda)^{n - n_0}\lambda_{\mathrm{term}}^{(n_0)}$,
converging geometrically to zero.

(iii) $K$-balance:
The multiplicative update~\eqref{eq:kterm_update} is balanced
when the fraction of updates with
$S_n < \tau_{\mathrm{safe}}$ equals
\begin{equation}\label{eq:p_star}
  p^{\ast} = \frac{-\log\kappa_{\downarrow}}
       {\log\kappa_{\uparrow} - \log\kappa_{\downarrow}}\,.
\end{equation}
When this fraction exceeds~$p^{\ast}$,
$K_{\mathrm{term}}$ grows toward~$K_{\max}$;
when it is below~$p^{\ast}$,
$K_{\mathrm{term}}$ decays toward~$K_{\min}$.
The balance point depends only on the multiplicative factors
$\kappa_\uparrow, \kappa_\downarrow$
and not on the policy or reward structure.

\textbf{Proof.}
\emph{Part~(i): Boundedness.}
For $K_{\mathrm{term}}$: the update~\eqref{eq:kterm_update}
applies $\min(K_{\max}, \cdot)$ in the increasing branch and
$\max(K_{\min}, \cdot)$ in the decreasing branch, so
$K^{(n+1)} \in [K_{\min}, K_{\max}]$ whenever
$K^{(n)} \in [K_{\min}, K_{\max}]$. Since the initialization
satisfies this, the bound holds for all~$n$ by induction.
The same argument applies to $\lambda_{\mathrm{term}}$ via the
projection to $[0, \lambda_{\max}]$
in~\eqref{eq:lambda_update}.

\emph{Part~(ii): $\lambda$-relaxation.}
When $\bar{\xi}_{\mathrm{term}}^{(n)} = 0$ for all
$n \geq n_0$, the violation branch
of~\eqref{eq:lambda_update} is never active, and the update
becomes $\lambda^{(n+1)} = (1-\delta_\lambda)\lambda^{(n)}$.
Unrolling the recursion from $n_0$:
$\lambda^{(n)} = (1-\delta_\lambda)^{n-n_0}\lambda^{(n_0)}$.
Since $\delta_\lambda \in (0,1)$, the factor
$(1-\delta_\lambda) \in (0,1)$ and
$(1-\delta_\lambda)^{n-n_0} \to 0$ geometrically as
$n \to \infty$. The rate of decay is
$-\log(1-\delta_\lambda) \approx \delta_\lambda$
for $\delta_\lambda \ll 1$.

\emph{Part~(iii): $K$-balance.}
Over $N$ updates in the interior of $[K_{\min}, K_{\max}]$,
suppose a fraction~$p$ have $S_n < \tau$ and
fraction~$(1{-}p)$ have $S_n \geq \tau$.
The net multiplicative factor is
$\kappa_\uparrow^{pN} \cdot \kappa_\downarrow^{(1-p)N}$.
This equals~$1$ (no net growth) when
$\kappa_\uparrow^{p} \cdot \kappa_\downarrow^{1-p} = 1$.
Taking logarithms:
$p\,\log\kappa_\uparrow + (1{-}p)\,\log\kappa_\downarrow = 0$.
Solving for~$p$:
$p\,\log\kappa_\uparrow = (1{-}p)\,|\log\kappa_\downarrow|$,
which gives~\eqref{eq:p_star}.
When $p > p^{\ast}$ the net factor exceeds~$1$ and $K_{\mathrm{term}}$
grows toward~$K_{\max}$; when $p < p^{\ast}$ it decays
toward~$K_{\min}$.
\hfill$\square$

\textbf{Remark~3} (Dual variable roles).
The $K_{\mathrm{term}}$ update~\eqref{eq:kterm_update} enforces the
chance constraint
$\Pr(\xi_{\mathrm{term}} = 0) \geq \tau_{\mathrm{safe}}/100$
through multiplicative adjustments in the log domain, with the
bounded range $[K_{\min},K_{\max}]$ preventing unbounded growth.
The $\lambda_{\mathrm{term}}$ update operates on a faster timescale
(additive steps proportional to violation magnitude). The two provide
complementary dual signals,
with $\lambda_{\mathrm{term}}$ targeting expected violation magnitude
and $K_{\mathrm{term}}$ targeting the probability of constraint
satisfaction, analogous in structure to two-timescale
dual ascent~\cite{c_borkar2008}, though formal
timescale-separation conditions are not verified here.

\section{VALIDATION ON TERMINALLY CONSTRAINED POWERTRAIN CONTROL}

The method is evaluated on a series-hybrid powertrain energy management
task with mixed continuous-discrete actions: binary engine on/off
(Categorical), continuous engine power, and ammonia-to-NO$_x$ ratio.
Each episode spans $T = 3{,}605$ steps. The terminal constraint
requires $|x_T^{\ast} - 0.55| \leq 0.002$ (SOC safe band
$[0.548,\, 0.552]$). The policy is a Transformer-encoder
actor~\cite{c_agarwal2023, c_vaswani2017} with two self-attention
layers, four heads, and a model dimension of~$64$, operating on a
sliding window of recent observations; no critic network is used.
Hyperparameters are listed in Table~\ref{tab:hyperparams}.

\begin{table}[h]
\vspace{-1.0em}
\centering
\caption{Hyperparameters for the series-hybrid powertrain task.}
\label{tab:hyperparams}
\vspace{-2pt}
\small
\begin{tabular}{@{}lc@{\hspace{8pt}}lc@{}}
\toprule
\textbf{Parameter} & \textbf{Value} & \textbf{Parameter} & \textbf{Value} \\
\midrule
Horizon $T$ & 3605 & Parallel envs $B$ & 48 \\
Policy updates & 800 & Batch size & 8192 \\
Clip $\epsilon$ & 0.10 & $\beta_{\mathrm{KL}}$ & 0.10 \\
$\lambda_{\mathrm{term}}$ (cap) & 350 & $\lambda_{\mathrm{tail}}$ (cap) & 70 \\
$\psi_{\mathrm{term}}$ & 0.50 & $\psi_{\mathrm{tail}}$ & 0.00 \\
$K_{\max}$ & 0.10 & $K_{\mathrm{center}}$ & 0.10 \\
$\kappa_\uparrow / \kappa_\downarrow$ & 1.01 / 0.995 & $\tau_{\mathrm{safe}}$ & 99\% \\
Learning rate & $10^{-4} \!\to\! 10^{-5}$ & LR schedule & Cosine \\
Feas.\ shift $s$ & 0.50 & Precision phase & Last 20\% \\
\bottomrule
\end{tabular}
\vspace{-4pt}
\end{table}

Four seeds (456, 3141, 350, 42) are trained with identical
hyperparameters. Performance is measured by the peak feasibility
rate and the sustained feasibility rate (mean over the last
$20\%$ of updates).
As baselines, dynamic programming (DP) with full model knowledge
yields a return of~$-226$, and a PPO-Lag baseline using the same
environment, horizon, and constraint definition is trained with
three seeds.

\subsection{Training Dynamics}

Figure~\ref{fig:training}(a) shows the shaped return for both
A-GRPO and PPO-Lag. A-GRPO improves monotonically as both cost
decreases and the penalty shrinks, while PPO-Lag remains
substantially lower throughout training.
Figures~\ref{fig:training}(b) and~(c) show the feasibility rate.
A-GRPO exhibits a three-phase pattern: exploration (updates 0--150,
near-zero feasibility), rapid improvement (150--400,
climbing to $60$--$80\%$), and refinement (400--800,
high but oscillatory feasibility).

The mean sustained feasibility rate across four A-GRPO seeds is
$75.4\% \pm 6.8\%$, with a mean peak of $95.8\%$ and a mean return
of $-234.3$ (within $3.7\%$ of the DP optimum; see
Fig.~\ref{fig:training}(a)). PPO-Lag achieves only
$27.4\% \pm 9.7\%$ mean sustained feasibility under the same
terminal constraint.
The best seed~(456) achieves $100\%$ peak, $84.0\%$ sustained,
and $-232.2$ return; the remaining seeds (42, 350, 3141) range from
$66.9\%$ to $79.7\%$ sustained.

\subsection{Dual Variable Analysis}

\begin{figure*}[t]
\centering
\includegraphics[width=\textwidth]{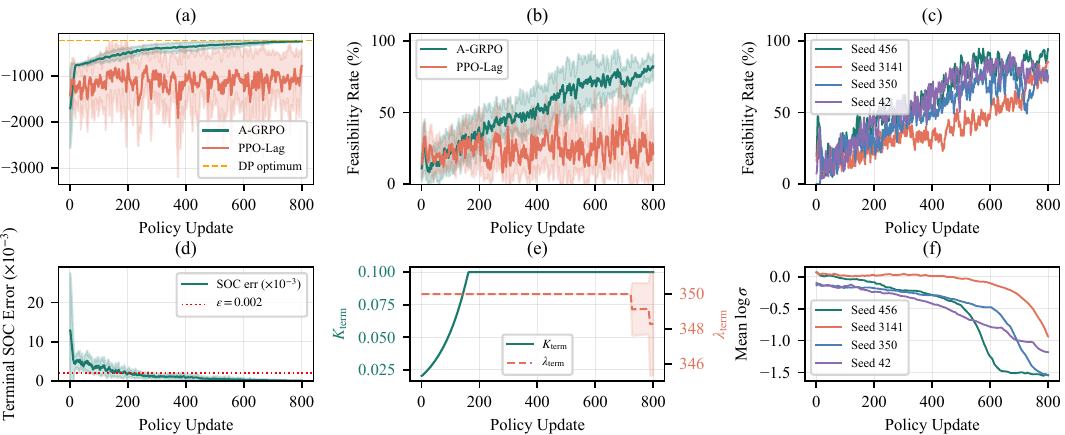}
\caption{Training dynamics.
(a)~Shaped return: A-GRPO (mean$\,\pm\,$std, 4 seeds) vs.\ PPO-Lag
(3 seeds); dashed line: DP optimum ($-226$).
(b)~Mean$\,\pm\,$std feasibility rate: A-GRPO vs.\ PPO-Lag.
(c)~Per-seed feasibility over 800 updates.
(d)~Terminal SOC error (mean$\,\pm\,$std, 4 seeds); dotted line: tolerance $\varepsilon = 0.002$.
(e)~Dual variables: $K_{\mathrm{term}}$ saturates at $K_{\max}$
by update~200; $\lambda_{\mathrm{term}}$ decays late for
high-feasibility seeds.
(f)~Policy logstd compression.}
\label{fig:training}
\vspace{-1.5em}
\end{figure*}

Figure~\ref{fig:training}(e) displays the evolution of the dual
variables $K_{\mathrm{term}}$ and $\lambda_{\mathrm{term}}$. For all
four seeds, $K_{\mathrm{term}}$ saturates at its cap
$K_{\max} = 0.10$ by approximately update~200 and remains there for
the duration of training. This behavior is consistent with
Theorem~2(iii): $\tau_{\mathrm{safe}} = 99\%$ serves as a
driving signal for $K_{\mathrm{term}}$ adaptation rather than an
achieved guarantee; since the batch feasibility rate never reaches
this target, $p \gg p^{\ast}$ in~\eqref{eq:p_star} and
$K_{\mathrm{term}}$ grows monotonically to its cap.
The cap is retained because larger ranking weights degrade
training (Table~\ref{tab:ablation_extra}).
Condition~\eqref{eq:kterm_condition} is conservative, guarding
against the worst-case return difference between feasible and
violating rollouts; in practice the Lagrangian penalty already
lowers the shaped returns of violating rollouts, so the separation
is achieved jointly by the return gap and the ranking bias,
consistent with the ablation results in Section~VI-C.
Early saturation gives the two mechanisms complementary roles:
the adaptive update establishes full violation pressure within
the first 200 updates, and policy variance compression drives
the subsequent feasibility gains (Figure~\ref{fig:training}(f)).
The target $\tau_{\mathrm{safe}} = 99\%$ maintains this pressure
throughout training, and the sustained rate of $75.4\%$ is set
by the degree of variance compression achieved on the tight
$\pm\varepsilon$ band.
The multiplier $\lambda_{\mathrm{term}}$ remains near
its cap of~$350$ throughout training.
Consistent with Theorem~2(ii), geometric decay activates only
for seed~456, which occasionally achieves $100\%$ batch feasibility;
other seeds maintain positive violations, keeping
$\lambda_{\mathrm{term}}$ saturated.

\subsection{Ablation Study}

\begin{figure}[t]
\centering
\includegraphics[width=\columnwidth]{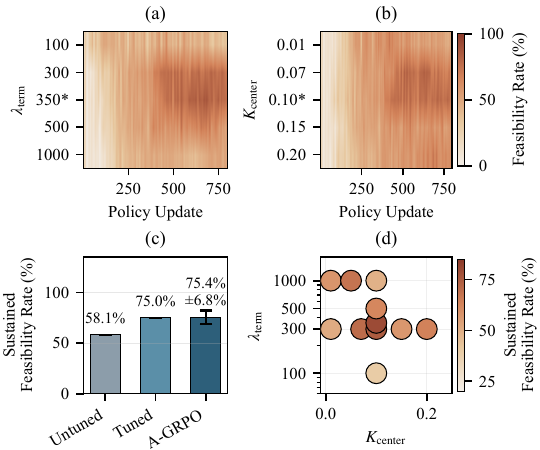}
\caption{Ablation study.
(a)~$\lambda_{\mathrm{term}}$ sweep heatmap with
$K_{\mathrm{center}} = 0.10$ fixed (seed~42); starred row denotes
the selected value ($\lambda_{\mathrm{term}} = 350$).
(b)~$K_{\mathrm{center}}$ sweep heatmap with
$\lambda_{\mathrm{term}} = 300$ fixed (seed~42); starred row denotes
$K_{\mathrm{center}} = 0.10$.
(c)~Sustained feasibility rate (last 20\%) for untuned ($\lambda_{\mathrm{term}} = 1000$,
$K_{\mathrm{center}} = 0.01$), tuned ($\lambda_{\mathrm{term}} = 300$, $K_{\mathrm{center}} = 0.10$),
and A-GRPO (4-seed mean$\,\pm\,$std).
(d)~$\lambda_{\mathrm{term}}$--$K_{\mathrm{center}}$ interaction
map: markers denote tested configurations, colored by sustained
feasibility rate; the joint optimum lies at $\lambda_{\mathrm{term}} = 350$,
$K_{\mathrm{center}} = 0.10$.}
\label{fig:ablation}
\vspace{-1.5em}
\end{figure}

Two sweeps assess the individual and joint contributions of the two
mechanisms that set the strength of the constraint signal, the
Lagrangian penalty ($\lambda_{\mathrm{term}}$) and the ranking
mechanism ($K_{\mathrm{center}}$), with results shown in
Figure~\ref{fig:ablation}; the remaining design factors are
summarized in Table~\ref{tab:ablation_extra}. All ablation runs use
seed~42.

In the Lagrangian sweep (Figure~\ref{fig:ablation}(a),
$K_{\mathrm{center}} = 0.10$ fixed), sweeping
$\lambda_{\mathrm{term}}$ over $\{100, 300, 350, 500, 1000\}$
reveals a non-monotone relationship with feasibility. The sustained
feasibility rate increases from $39.8\%$ at $\lambda_{\mathrm{term}} = 100$ to a peak of
$79.7\%$ at $\lambda_{\mathrm{term}} = 350$, then declines to $49.2\%$ at
$\lambda_{\mathrm{term}} = 1000$. Under-penalization ($\lambda_{\mathrm{term}} = 100$) provides
insufficient constraint signal for the policy to learn terminal
targeting, while over-penalization ($\lambda_{\mathrm{term}} = 1000$) introduces
large gradient variance through the terminal reward, destabilizing
training.

In the ranking sweep (Figure~\ref{fig:ablation}(b),
$\lambda_{\mathrm{term}} = 300$ fixed), sweeping
$K_{\mathrm{center}}$ over $\{0.01, 0.07, 0.10, 0.15, 0.20\}$
similarly shows a peaked profile. The sustained rate rises from
$51.1\%$ at $K_{\mathrm{center}} = 0.01$ to $75.0\%$ at
$K_{\mathrm{center}} = 0.10$, then drops to $57.3\%$ at
$K_{\mathrm{center}} = 0.15$. Too-small $K_{\mathrm{center}}$
fails to differentiate near-feasible from far-feasible trajectories
within the batch, whereas too-large $K_{\mathrm{center}}$ dominates
the advantage signal and prevents the policy from compressing its
logstd, which is necessary for precise terminal targeting.

The default configuration ($\lambda_{\mathrm{term}} = 1000$,
$K_{\mathrm{center}} = 0.01$) achieves only $58.1\%$ sustained.
The optimized A-GRPO ($\lambda_{\mathrm{term}} = 350$, $K_{\mathrm{center}} = 0.10$)
improves this to $79.7\%$ ($+21.6$ pp). Neither reducing $\lambda_{\mathrm{term}}$
alone nor increasing $K_{\mathrm{center}}$ alone recovers this
performance. The improvement therefore follows from the interaction
of the ranking weight with the penalty magnitude, the
complementarity formalized in Theorem~1.
Figure~\ref{fig:ablation}(d) shows the joint interaction surface.

Table~\ref{tab:ablation_extra} summarizes seven additional factors.
Increasing $K_{\max}$ beyond $0.10$ monotonically degrades
performance, as large ranking weights distort advantages and impede
policy compression. A ramp rate of $\kappa_\uparrow = 1.03$
drives $K_{\mathrm{term}}$ to its cap too quickly, causing
destabilizing KL divergence growth; only $\kappa_\uparrow = 1.01$
remains stable, consistent with the gentle update dynamics
described in Remark~3. The precision phase, which applies
an entropy penalty during the final $20\%$ of training, yields
$+4$--$5\%$ sustained improvement by encouraging logstd compression.
Both training duration and environment count exhibit non-monotonic
sensitivity: extending beyond $800$ updates causes
over-compression of the policy variance (up to $47$ percentage
points of loss), while the environment count peaks at~$48$, since
fewer environments starve rollout diversity and more inflate policy
variance. Increasing $\beta_{\mathrm{KL}}$ past $0.10$ slows
learning without improving feasibility. A feasibility shift of
$s = 1.0$ is comparable to the standard deviation of the normalized
advantage, so the binary feasibility label dominates the within-batch
ordering and the policy collapses; $s = 0.50$ keeps the ordering
responsive to the return as well.

\begin{table}[h]
\vspace{-1.0em}
\centering
\caption{Sensitivity to additional design factors (\textbf{bold} = selected value).}
\vspace{-1.0em}
\label{tab:ablation_extra}
\footnotesize
\setlength{\tabcolsep}{3pt}
\begin{tabularx}{\columnwidth}{@{} p{0.65in} p{1.15in} X @{}}
\toprule
Factor & Tested & Finding \\
\midrule
$K_{\max}$  & \textbf{0.10}, 0.5, 1.0, 2.0
  & Higher monotonically worse \\
$\kappa_\uparrow$ & \textbf{1.01}, 1.03
  & 1.03 $\Rightarrow$ KL death \\
Precision & \textbf{ON}, OFF
  & $+4$--$5\%$ sustained \\
Updates & \textbf{800}, 1000, 1200
  & Extended over-compresses \\
Env count & 10, \textbf{48}, 96, 256
  & 10: collapse; 256: inflation \\
$\beta_{\mathrm{KL}}$ & \textbf{0.10}, 0.15, 0.20
  & Higher slows learning \\
Shift $s$ & \textbf{0.50}, 1.0
  & 1.0 $\Rightarrow$ policy collapse \\
\bottomrule
\end{tabularx}
\vspace{-1.5em}
\end{table}

\subsection{Policy Compression and Terminal Behavior}

Figure~\ref{fig:training}(f) shows monotonic logstd compression
across seeds. Seed~456 reaches logstd~$\approx -1.5$;
other seeds plateau between $-0.8$ and $-1.5$. Per-timestep
normalization (Proposition~1) keeps advantage signals well-scaled
during this compression.

Figure~\ref{fig:training}(d) shows the terminal SOC error dropping
below $\varepsilon = 0.002$ by update~200 and continuing to
$\sim\!10^{-5}$. The residual infeasible rollouts visible in
Figures~\ref{fig:training}(b)--(c) are individual trajectories,
not batch-mean violations, confirming that the per-trajectory ranking
(Theorem~1) provides finer differentiation than the Lagrangian's
batch-mean signal.
The number of such rollouts fluctuates from batch to batch as
sampled terminal states land on either side of the tight
$\pm\varepsilon$ band, which produces the late-phase oscillation
of the feasibility rate; accordingly, seeds with stronger variance
compression sustain higher feasibility
(Figure~\ref{fig:training}(f)).

\section{CONCLUSION}

The boundedness and feasibility properties of Advantage-Ranked Group
Relative Policy Optimization (A-GRPO), a ranking-augmented,
critic-free policy gradient method employing a Transformer-encoder
actor for terminally constrained control, have been analyzed in
this paper. Per-timestep advantage normalization bounds variance
at every timestep (Proposition~1), the ranked advantage separates
feasible from violating trajectories and biases the policy gradient
toward constraint satisfaction (Theorem~1), and the adaptive dual
variables remain bounded with a drift-balance property that acts
as a feedback mechanism for the chance constraint (Theorem~2).

Experiments on a $3{,}605$-step powertrain energy management task
confirmed these results: the combined ranking-Lagrangian mechanism
achieved $75.4\% \pm 6.8\%$ mean sustained feasibility across four
seeds with a mean return within $3.7\%$ of the dynamic programming
optimum, outperforming a PPO-Lag baseline ($27.4\%$
sustained) and either component in isolation. The observed training
dynamics were consistent with the theoretical predictions.
Future work will investigate adaptive variance scheduling to
reduce inter-seed sensitivity
and extensions to multi-constraint settings where several terminal
conditions must be satisfied simultaneously.

\addtolength{\textheight}{0cm}   



\begin{thebibliography}{99}

\bibitem{c_egan2023}
D.~Egan, Q.~Zhu, and R.~Prucka,
``A review of reinforcement learning-based powertrain controllers:
Effects of agent selection for mixed-continuity control and reward formulation,''
\textit{Energies}, vol.~16, no.~8, p.~3450, 2023.

\bibitem{c_rolando2024}
L.~Rolando, N.~Campanelli, L.~Tresca, L.~Pulvirenti, and F.~Millo,
``Development of a soft actor--critic reinforcement learning algorithm for the energy management of a hybrid electric vehicle,''
\textit{SAE Int. J. Advances and Current Practices in Mobility}, vol.~7, pp.~1140--1151, 2024.

\bibitem{c_altman1999}
E.~Altman,
\textit{Constrained Markov Decision Processes}.
Boca Raton, FL: Chapman and Hall/CRC, 1999.

\bibitem{c_achiam2017}
J.~Achiam, D.~Held, A.~Tamar, and P.~Abbeel,
``Constrained policy optimization,''
in \textit{Proc. Int. Conf. Machine Learning (ICML)},
pp.~22--31, PMLR, 2017.

\bibitem{c_ding2022}
D.~Ding, K.~Zhang, J.~Duan, T.~Ba\c{s}ar, and M.~R.~Jovanovi\'{c},
``Convergence and sample complexity of natural policy gradient primal-dual methods for constrained MDPs,''
\textit{arXiv preprint arXiv:2206.02346}, 2022.

\bibitem{c_paternain2019}
S.~Paternain, L.~F.~O.~Chamon, M.~Calvo-Fullana, and A.~Ribeiro,
``Constrained reinforcement learning has zero duality gap,''
in \textit{Advances in Neural Information Processing Systems}, vol.~32, pp.~7553--7563, 2019.

\bibitem{c_stooke2020}
A.~Stooke, J.~Achiam, and P.~Abbeel,
``Responsive safety in reinforcement learning by PID Lagrangian methods,''
in \textit{Proc. Int. Conf. Machine Learning (ICML)},
pp.~9133--9143, PMLR, 2020.

\bibitem{c_garcia2015}
J.~Garc\'{i}a and F.~Fern\'{a}ndez,
``A comprehensive survey on safe reinforcement learning,''
\textit{J. Machine Learning Research}, vol.~16, no.~1, pp.~1437--1480, 2015.

\bibitem{c_chow2017}
Y.~Chow, M.~Ghavamzadeh, L.~Janson, and M.~Pavone,
``Risk-constrained reinforcement learning with percentile risk criteria,''
\textit{J. Machine Learning Research}, vol.~18, no.~1, pp.~1--51, 2017.

\bibitem{c_berkenkamp2017}
F.~Berkenkamp, M.~Turchetta, A.~Schoellig, and A.~Krause,
``Safe model-based reinforcement learning with stability guarantees,''
in \textit{Advances in Neural Information Processing Systems}, vol.~30, pp.~908--918, 2017.

\bibitem{c_tessler2019}
C.~Tessler, D.~J.~Mankowitz, and S.~Mannor,
``Reward constrained policy optimization,''
in \textit{Proc. Int. Conf. Learning Representations (ICLR)}, 2019.

\bibitem{c_thananjeyan2021}
B.~Thananjeyan, A.~Balakrishna, S.~Nair, M.~Luo, K.~Srinivasan,
M.~Hwang, and K.~Goldberg,
``Recovery RL: Safe reinforcement learning with learned recovery zones,''
\textit{IEEE Robotics and Automation Letters}, vol.~6, no.~3, pp.~4915--4922, 2021.

\bibitem{c_schulman2017}
J.~Schulman, F.~Wolski, P.~Dhariwal, A.~Radford, and O.~Klimov,
``Proximal policy optimization algorithms,''
\textit{arXiv preprint arXiv:1707.06347}, 2017.

\bibitem{c_ray2019}
A.~Ray, J.~Achiam, and D.~Amodei,
``Benchmarking safe exploration in deep reinforcement learning,''
\textit{arXiv preprint arXiv:1910.01708}, 2019.

\bibitem{c_jayant2022}
A.~K.~Jayant and S.~Bhatnagar,
``Model-based safe deep reinforcement learning via a constrained proximal policy optimization algorithm,''
in \textit{Advances in Neural Information Processing Systems}, vol.~35, pp.~24432--24445, 2022.

\bibitem{c_haarnoja2018}
T.~Haarnoja, A.~Zhou, P.~Abbeel, and S.~Levine,
``Soft actor-critic: Off-policy maximum entropy deep reinforcement learning with a stochastic actor,''
in \textit{Proc. Int. Conf. Machine Learning (ICML)}, pp.~1861--1870, PMLR, 2018.

\bibitem{c_fujimoto2018}
S.~Fujimoto, H.~Hoof, and D.~Meger,
``Addressing function approximation error in actor-critic methods,''
in \textit{Proc. Int. Conf. Machine Learning (ICML)}, pp.~1587--1596, PMLR, 2018.

\bibitem{c_agarwal2023}
P.~Agarwal, A.~A.~Rahman, P.-L.~St-Charles, S.~J.~D.~Prince, and S.~E.~Kahou,
``Transformers in reinforcement learning: A survey,''
\textit{arXiv preprint arXiv:2307.05979}, 2023.

\bibitem{c_yuan2024}
W.~Yuan, J.~Chen, S.~Chen, D.~Feng, Z.~Hu, P.~Li, and W.~Zhao,
``Transformer in reinforcement learning for decision-making: A survey,''
\textit{Frontiers of Information Technology \& Electronic Engineering},
vol.~25, no.~6, pp.~763--790, 2024.

\bibitem{c_shao2024}
Z.~Shao, P.~Wang, Q.~Zhu, R.~Xu, J.~Song, X.~Bi, et~al.,
``DeepSeekMath: Pushing the limits of mathematical reasoning in open language models,''
\textit{arXiv preprint arXiv:2402.03300}, 2024.

\bibitem{c_guo2025}
D.~Guo, D.~Yang, H.~Zhang, J.~Song, R.~Zhang, R.~Xu, et~al.,
``DeepSeek-R1: Incentivizing reasoning capability in LLMs via reinforcement learning,''
\textit{arXiv preprint arXiv:2501.12948}, 2025.

\bibitem{c_lin2025}
Z.~Lin, M.~Lin, Y.~Xie, and R.~Ji,
``CPPO: Accelerating the training of group relative policy optimization-based reasoning models,''
\textit{arXiv preprint arXiv:2503.22342}, 2025.

\bibitem{c_li2025}
C.~Li, N.~Liu, and K.~Yang,
``Adaptive group policy optimization: Towards stable training and token-efficient reasoning,''
\textit{arXiv preprint arXiv:2503.15952}, 2025.

\bibitem{c_guo2025b}
Y.~Guo, W.~Deng, Z.~Cheng, and X.~Tang,
``G$^2$RPO-A: Guided group relative policy optimization with adaptive guidance,''
\textit{arXiv preprint arXiv:2508.13023}, 2025.

\bibitem{c_khanda2025}
R.~Khanda, M.~Baqar, S.~Chakrabarti, and S.~Changdar,
``Extending group relative policy optimization to continuous control:
A theoretical framework for robotic reinforcement learning,''
\textit{arXiv preprint arXiv:2507.19555}, 2025.

\bibitem{c_sane2025}
S.~Sane,
``Hybrid group relative policy optimization: A multi-sample approach to enhancing policy optimization,''
\textit{arXiv preprint arXiv:2502.01652}, 2025.

\bibitem{c_chen2019pt}
I.~M.~Chen, C.~Zhao, and C.~Y.~Chan,
``A deep reinforcement learning-based approach to intelligent powertrain control for automated vehicles,''
in \textit{Proc. IEEE Intelligent Transportation Systems Conference (ITSC)},
pp.~2620--2625, 2019.

\bibitem{c_rownak2026}
M.~R.~Rownak, W.~Jaleel, A.~Hanif, M.~Q.~Fahim, D.~D.~Le, H.~Anwar,
M.~Nelson, and Q.~Ahmed,
``Physics-aware deep reinforcement learning for energy and aging management
in electrified powertrains,''
\textit{IEEE Trans. Transportation Electrification}, early access, 2026,
doi: 10.1109/TTE.2026.3691278.

\bibitem{c_sutton2018}
R.~S.~Sutton and A.~G.~Barto,
\textit{Reinforcement Learning: An Introduction}, 2nd~ed.
Cambridge, MA: MIT Press, 2018.

\bibitem{c_bertsekas1999}
D.~P.~Bertsekas,
\textit{Nonlinear Programming}, 2nd~ed.
Belmont, MA: Athena Scientific, 1999.

\bibitem{c_williams1992}
R.~J.~Williams,
``Simple statistical gradient-following algorithms for connectionist reinforcement learning,''
\textit{Machine Learning}, vol.~8, no.~3--4, pp.~229--256, 1992.

\bibitem{c_sutton2000}
R.~S.~Sutton, D.~McAllester, S.~Singh, and Y.~Mansour,
``Policy gradient methods for reinforcement learning with function approximation,''
in \textit{Advances in Neural Information Processing Systems}, vol.~12,
pp.~1057--1063, 2000.

\bibitem{c_vaswani2017}
A.~Vaswani, N.~Shazeer, N.~Parmar, J.~Uszkoreit, L.~Jones,
A.~N.~Gomez, \L.~Kaiser, and I.~Polosukhin,
``Attention is all you need,''
in \textit{Advances in Neural Information Processing Systems}, vol.~30, pp.~5998--6008, 2017.

\bibitem{c_borkar2008}
V.~S.~Borkar,
\textit{Stochastic Approximation: A Dynamical Systems Viewpoint}.
Cambridge, UK: Cambridge University Press, 2008.

\bibitem{c_nedic2009}
A.~Nedi\'{c} and A.~Ozdaglar,
``Approximate primal solutions and rate analysis for dual subgradient methods,''
\textit{SIAM J. Optimization}, vol.~19, no.~4, pp.~1757--1780, 2009.

\bibitem{c_anthropic2025}
Anthropic,
``Claude,'' 2025.
[Online]. Available: \url{https://www.anthropic.com/claude}

\end{thebibliography}
\end{document}